\documentclass[apj]{emulateapj}
\usepackage{graphics}
\usepackage{natbib}
\newcommand{\as}{$''$}

\newcommand{\nh}{$N_{\rm{H}}$}
\newcommand{\nustar}{\textit{NuSTAR}}
\newcommand{\integral}{\textit{INTEGRAL}}
\newcommand{\chandra}{\textit{Chandra}}
\newcommand{\rxte}{\textit{RXTE}}
\newcommand{\fermi}{\textit{Fermi}}
\newcommand{\psr}{J1811-1925}
\newcommand{\gpsr}{G11.2-0.3}
\newcommand{\degree}{$^\circ$}
\newcommand{\src}{G11.2-0.3}

\begin{document}

\title{NuSTAR observations of \gpsr.}
\author{K. K. Madsen$^1$, C. L. Fryer$^2$, B. W. Grefenstette$^1$, L. A. Lopez$^3$, S. Reynolds$^4$, A. Zoglauer$^5$ }

\affiliation{
$^1$ Cahill Center for Astronomy and Astrophysics, California Institute of Technology, Pasadena, CA 91125, USA\\
$^2$ CCS-2, Los Alamos National Laboratory, Los Alamos, NM 87545, USA\\
$^3$ The Ohio State University, Columbus, OH 43210, USA\\
$^4$ Physics Department, NC State University, Raleigh, NC 27695, USA\\
$^5$ Space Sciences Laboratory, University of California, Berkeley, CA 94720, USA\\
}

\begin{abstract}
We present in this paper the hard X-ray view of the pulsar wind nebula in \src\ and its central pulsar PSR \psr\ as seen by \nustar. We complement the data with \chandra\ for a more complete picture and confirm the existence of a hard, power-law component in the shell with photon index $\Gamma = 2.1 \pm 0.1$, which we attribute to synchrotron emission. Our imaging observations of the shell show a slightly smaller radius at higher energies, consistent with {\sl Chandra} results, and we find shrinkage as a function of increased energy along the jet direction, indicating that the electron outflow in the PWN may be simpler than that seen in other young PWNe.  Combining \nustar\ with \integral, we find that the pulsar spectrum can be fit by a power-law with $\Gamma=1.32 \pm 0.07$ up to 300~keV without evidence of curvature.
\end{abstract}


\section{Introduction}

According to conventional ideas, the young remnant of a core-collapse
supernova (CCSN) ought to consist of a shell emitting brightly in
radio synchrotron emission and thermal X-rays, containing a pulsar and
pulsar-wind nebula (PWN).  The Galactic pulsar birthrate \citep[of
  order 1 -- 2 per century;][]{Vranesevic04}, in combination with
standard estimates of the Galactic CCSN rate of 2 -- 3 per century
\citep[e.g.,][]{Tammann94} requires that a large fraction of CCSNe
should produce pulsars, and any self-respecting pulsar ought to
inflate a bright synchrotron nebula.  Furthermore, something like 80\%
of supernovae should be CCSNe \citep{Tammann94}.  The remnants of
recent supernovae in our Galaxy fail significantly to live up to this
expectation.  The well-documented historical or quasi-historical
supernovae of the past two millenia include five (likely) Type Ia SNe:
G1.9+0.2 \citep[ca.~1900 CE;][]{Reynolds08}, Kepler (SN 1604), Tycho
(SN 1572), SN 1006, and RCW 86 \citep[SN 185;][]{Williams08}, and two
atypical CCSN remnants: Cas A, with a central non-pulsing neutron star
\citep{pavlov00}, and the Crab, whose absence of any kind of external shell is a
continuing embarrassment \citep[unless the ``shell'' is emission at
  the edge of the synchrotron nebula;][]{Hester2008}.

However, several other Galactic remnants are clearly quite young,
though without as clear age documentation.  The youngest of all known
CCSN remnants containing a PWN is Kes 75 (G29.7$-$0.3), with an age estimated from
expansion of $480 \pm 50$ years \citep{Reynolds2018}.  It has a very
asymmetric partial shell surrounding a bright PWN.  The next youngest,
once associated with a claimed historical SN in 386 AD, but now known
to suffer too much extinction to have been a naked-eye supernova, is
G11.2--0.3, with an expansion age of 1400 -- 2400 years
\citep{Borkowski2016}, which has all the expected components of a
young CCSN: distinct, fairly symmetric shell, and bright PWN with a
jet/torus structure as often seen \citep{ng04} containing a 65 ms
pulsar.

In principle, the youngest objects should provide the most information
about their birth events, their immediate surroundings, and the nature
of the freshly created pulsars.  Much of that information is
accessible through study of X-ray emission of a few keV energy:
thermal X-ray emission from ejecta and swept-up ambient medium, and
the spectrum and morphology of the non-thermal emission from the PWN. The
pulsar itself may be detectable in X-rays.  An analysis of early {\sl
  Chandra} observations of \src\ \citep{Roberts2003} found possible evidence for
a hard, perhaps non-thermal, spectral component in the shell, while
characterizing the PWN spectrum between 1 and 10 keV.  But {\sl
  Chandra}'s bandpass, while ideal for thermal emission, is not wide
enough to allow firm conclusions to be drawn on the spectral slope
(and spatial structure) in the PWN, or to clearly separate any
non-thermal emission from the shell's thermal emission.  These analyses
become much more straightforward at higher energies, in the
range ideally suited to {\sl NuSTAR.}

An in-depth analysis of a 400 ks {\sl Chandra} observation of
\src\ found a number of puzzles \citep{Borkowski2016}.  The shell spectrum indicates a large swept-up mass.  Expansion into
a uniform medium, or into a steady spherical wind ($\rho \propto
r^{-2}$), are ruled out by evolutionary considerations.  But a
combination of morphological and spectral information on the shell
interior implies that the reverse shock has already returned to the
center of the remnant, confining and compressing the PWN.  The PWN
itself shows no significant spectral steepening as one moves away from
the pulsar, unlike most other PWNe \citep[e.g.,][]{Bocchino2001}.  This
fact has implications for the nature of particle transport in PWNe.

While many of these questions require examination of the thermal
emission, non-thermal emission at higher energies can address issues of
particle acceleration in both the shell and PWN and of PWN
evolution. If the shell of \src\ has no associated synchrotron X-ray
emission, \src\ will be alone among remnants less than a few thousand
years old in this property.  \cite{Borkowski2016} found blast-wave
velocities from direct expansion of 700 -- 1200 km s$^{-1}$, fast
enough to allow electron acceleration to X-ray-emitting energies.  The
confirmation and spectral characterization of such emission is
important for the study of shock acceleration.  The PWN is also one of
the youngest known, and its spectral properties above the {\sl
  Chandra} band are important for the study of particle acceleration
in relativistic shocks and transport into the PWN interior.

Power to the PWN is provided by the central rotation powered pulsar (PSR). How exactly the pulsar manages to produce its wind and how this wind becomes particle dominated are questions that are still unanswered, but observational properties of the engine can shed light on the problem by providing clues to the geometry of particle acceleration in the magnetospheres. \psr\ is a radio-quiet, $\sim$65~ms, high-magnetic field pulsar with a field strength of $B \sim 10^{12}$ G and an estimated rotational kinetic energy loss of $\dot{E}\sim 6 \times 10^{36}$ ergs s$^{-1}$ \citep{Tori1999}. It was discovered in soft X-rays by \textit{ASCA} \citep{Torii1997} and in the soft (20 -- 300~keV) $\gamma$-ray band by \integral/IBIS \citep{Dean2008}. Many PWN have proven themselves to be effective accelerators, and due to proximity, it was postulated whether \psr\ could be related to the nearby TeV source HESS~J1809-193, but the association was deemed unlikely due to the distance from the TeV emitter and the fact that the jet of \psr\ is not pointed towards HESS~J1809-193, in which case it becomes hard to explain how the particles are propagating to the target. It remains undetected in radio \citep{Crawford1998} and in the GeV by the \textit{Fermi} LAT \citep{Acero2016}, which makes the X-ray band the only accessible for study. 

The hard x-ray properties of the pulsar have been previously studied with \rxte\ \citep{Roberts2004}, where it was seen that the pulse profile maintained its sinusoidal shape up to 90\,keV, and the pulsed spectrum was measured in the PCA (2.5 -- 30 \,keV) to be a power-law with slope $\Gamma = 1.16 \pm 0.2$. Later,  the data from \rxte\ was combined with \integral\ \citep{Kuiper2015}, confirming pulsations up to 135\,keV, and the spectrum of the remnant + pulsar above 20\,keV to  be consistent with a power-law of $\Gamma = 1.61 \pm 0.15$ .

In this paper we undertook a detailed study of \src\ with
\nustar\, to examine the pulsar, PWN, and shell.  For the pulsar, we
examined the pulse profiles as a function of energy and the spectrum
of pulsations. For the PWN, we examined the integrated spectrum and
energy-dependent morphology. Finally, for the shell, we attempted to confirm the presence of
non-thermal emission in the shell and to study it if confirmed.



\begin{figure}
\begin{center}
\includegraphics[width=0.5\textwidth]{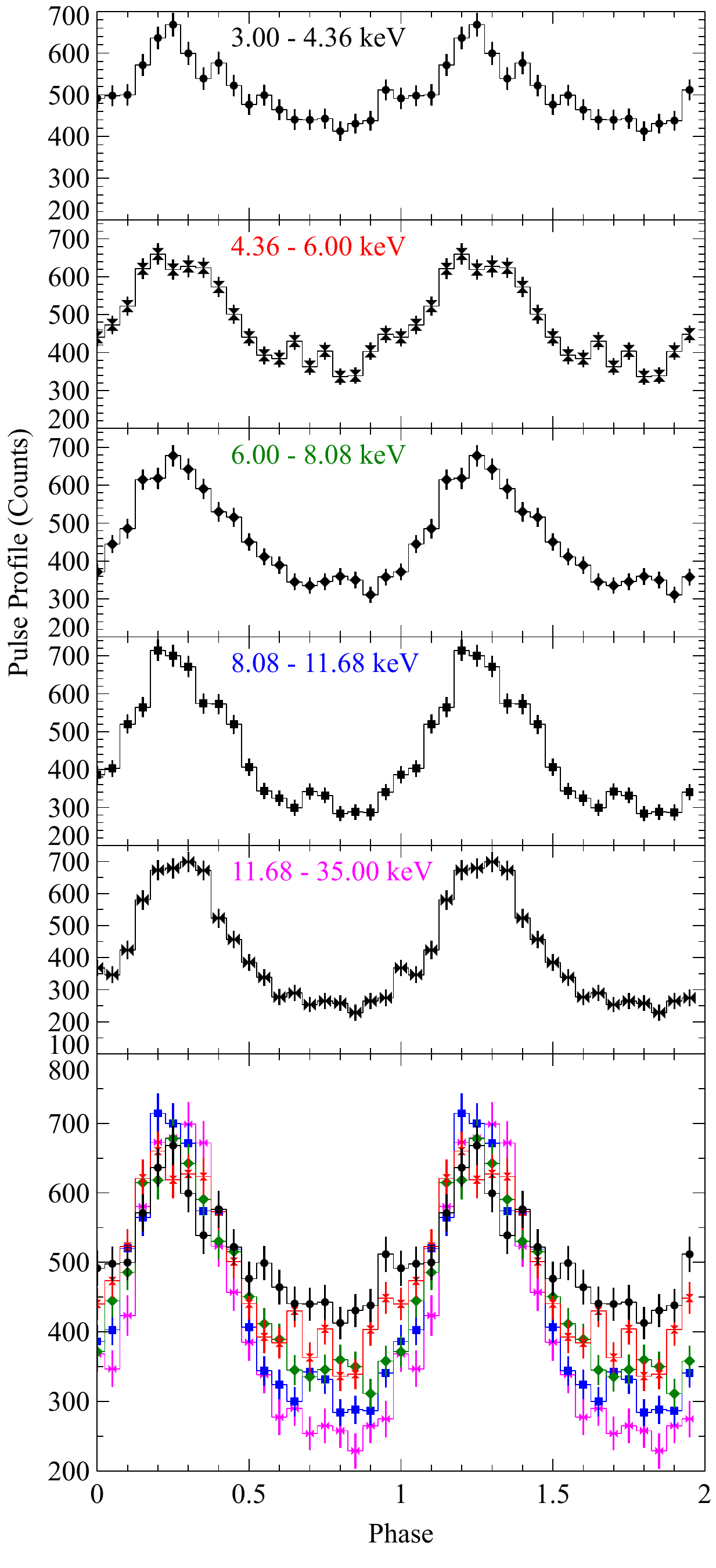}
\end{center}
\caption{Pulse profiles in different energy bands. Energy bins were selected to have an equal number of net counts ($\sim$7800 counts). We define the off-pulse period as phase 0.6 -- 1.0. The bottom panel compares the pulse profiles in the five energy bins. Though the integrated flux at pulse peak is the same across these energy bins, the off-peak flux decreases with energy.}
\label{pulseprofile}
\end{figure}

\begin{figure}
\begin{center}
\includegraphics[width=0.5\textwidth]{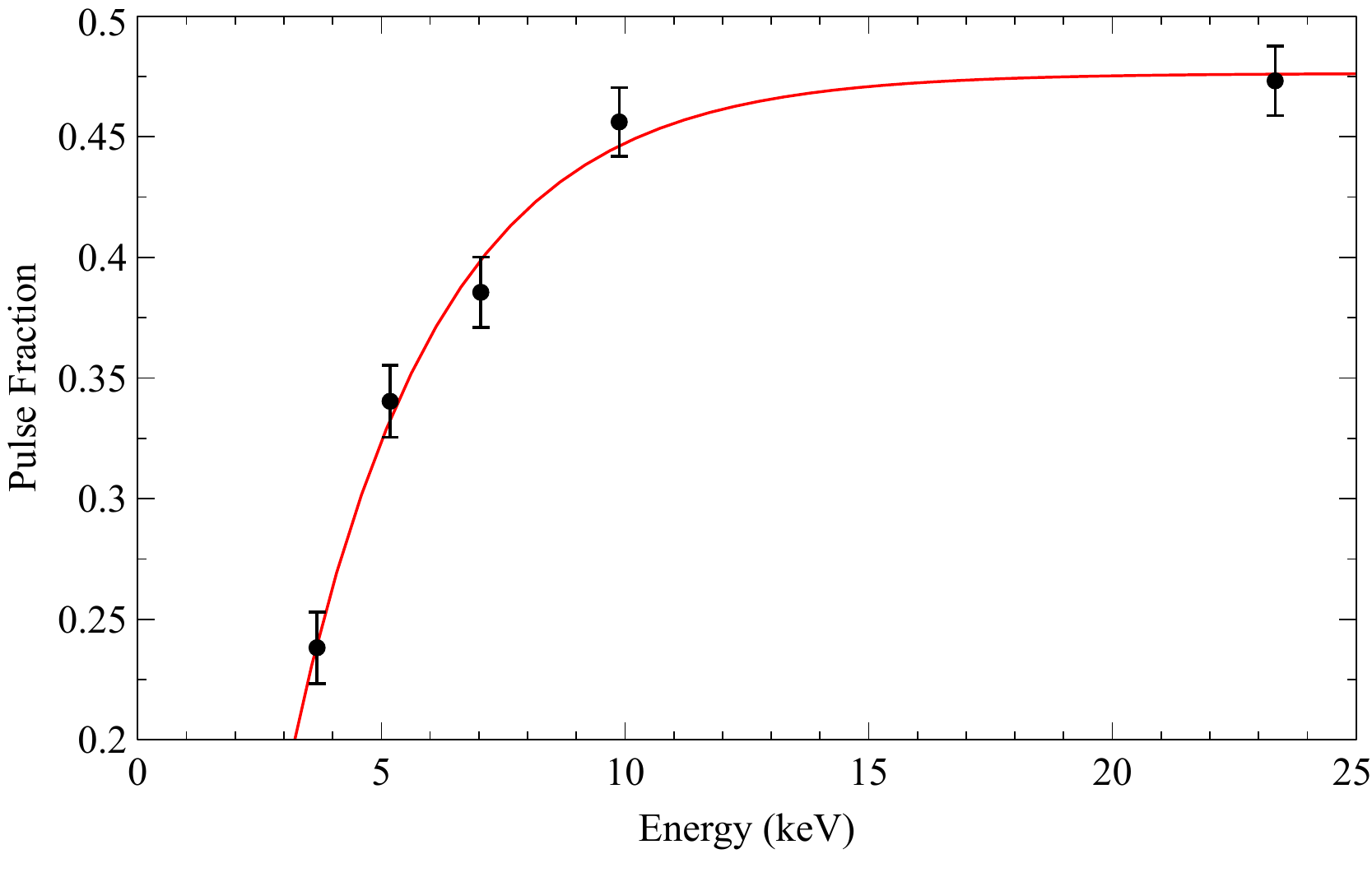}
\end{center}
\caption{Pulse fraction as a function of energy fitted with an exponential function. As the pulse fraction is representative of the contribution of the pulse to the steady PWN component, the curves shows that the PWN flux gets fainter with respect to the pulsed component with increasing energy.}
\label{pulsefraction}
\end{figure}

\section{Observations and Data Reduction}
We use in this paper data from \nustar\ \citep{Harrison2013}, \chandra\ \citep{Weisskopf2002}, and \integral\ \citep{Ubertini2003}. The \nustar\ data was taken from June 23 to June 26, 2016 for a total on target exposure time of 89ks after filtering away high SAA regions. The data was reduced using \texttt{nustardas\_0.9Jun15\_v1.5.1} and CALDB version \texttt{20160606}. Several stray-light regions appear close to the source, as well as a transient source located at RA=272:49:16.93 and Dec=-19:28:23.16, which appeared briefly between 2016 June 24 at 22:57:00 UTC and 2016 June 25 at 00:20:00 UTC, but the source itself is clear of contamination, and a clean background region could be obtained adjacent to the source. The details of each extracted spectrum will be visited in the relevant sections. \nustar\ flies two co-aligned telescopes with two identical detector focal planes that we will refer to as FPMA and FPMB.

{\sl Chandra} data used here were obtained in five segments between May 5 and September 9, 2013 for a total effective exposure of 388 ks after screening, as described in \cite{Borkowski2016}.  Data were obtained in Very Faint mode, and reprocessed with CIAO v4.6 and CALDB v4.6.3.  Screening for periods of high particle background was performed.  The five observations were aligned as described in \cite{Borkowski2016}.  For spectral analysis, the background had to be obtained from the \chandra\ blank fields, available through the \chandra\ CALDB\footnote{See http://cxc.harvard.edu/ciao/download/caldb.html}. This was necessary, because the source contaminates the entire S3 CCD chip, most likely due to dust scattering, which will be significant at column densities above $1 \times 10^{22}$\,atoms\,cm$^{-2}$. 

We obtained the \integral\ ISGRI/IBIS data and responses on PSR J1811+1925 (we note that in the catalog it is labeled as PSR J1811+1926) from the \integral\ General Reference Catalogue v.41 \citep{Ebisawa2003}\footnote{See https://www.isdc.unige.ch/integral/science/catalogue}.

\section{Analysis}
We will first present the analysis of the pulse profile, \S\ref{sec:pulseprofile}, then the analysis of the geometrical properties of the remnant in \S\ref{sec:imaging}, both of which were performed with \nustar\ data only. The spectral analysis section will address separately in order; the full remnant broad band spectrum, \S\ref{sec:broad}; the nebula spectrum, \S\ref{sec:nebula}; the pulsed spectrum, \S\ref{sec:pulse}; and finally the broadband spectral energy distribution (SED) from 1 -- 300~keV, \S\ref{sec:sed}.

\subsection{Pulse profile}\label{sec:pulseprofile}
We applied barycenter corrections to the event file using the the \nustar\ clock-correction file version 32 (or newer) at the position of the pulsar as given by \chandra. We extracted source counts from a 123\as\ radius circular region (corresponding to 50 pixels) and added FPMA and FPMB counts together. We applied the ephemeris provided by \citet{Smith2008}\footnote{https://fermi.gsfc.nasa.gov/ssc/data/access/lat/ephems/}, which was obtained from \rxte, and folded the lightcurve, but did not recover the pulsations. We then used HENDRICS \citep{Bachetti2015}, built on Stingray \citep{Huppenkothen2016}, to find a new local solution and PINT to calculate the errors \citep{Luo2015}. We obtain a frequency of $\nu = 15.4564269(1)$ Hz and $\dot\nu = -8.(2)e-12$ Hz s$^{-1}$ at the epoch $T_0$= 57563 MJD.
We then chose 5 different energy bins, selected to have an equal amount of counts in each bin after background subtraction to get the pulse profiles shown in Figure \ref{pulseprofile}. The energy bins each contain $\sim 7800$ counts and are: 3.00 -- 4.36, 4.36 -- 6.00, 6.00 -- 8.08, 8.08 -- 11.68, and 11.68 -- 35.00\,keV. The integrated flux at pulse peak remains approximately constant throughout the five bins, while the off-pulse flux decreases, causing the pulse to broaden. We calculate the pulse fraction as $PF = (F_{max}-F_{min})/(F_{max}+F_{min})$, where $F_{min}$ is the minimum flux in the pulse profile and $F_{max}$ the maximum flux in the profile. The resulting curve, shown in Figure \ref{pulsefraction}, can be fit with an exponential that flattens above 25~keV. Since the pulse fraction is a measure of the ratio of the pulse to the steady PWN component, the curve shows the PWN flux growing fainter with respect to the pulsed component as a function of increasing energy.

\begin{figure}
\begin{center}
\includegraphics[width=0.5\textwidth]{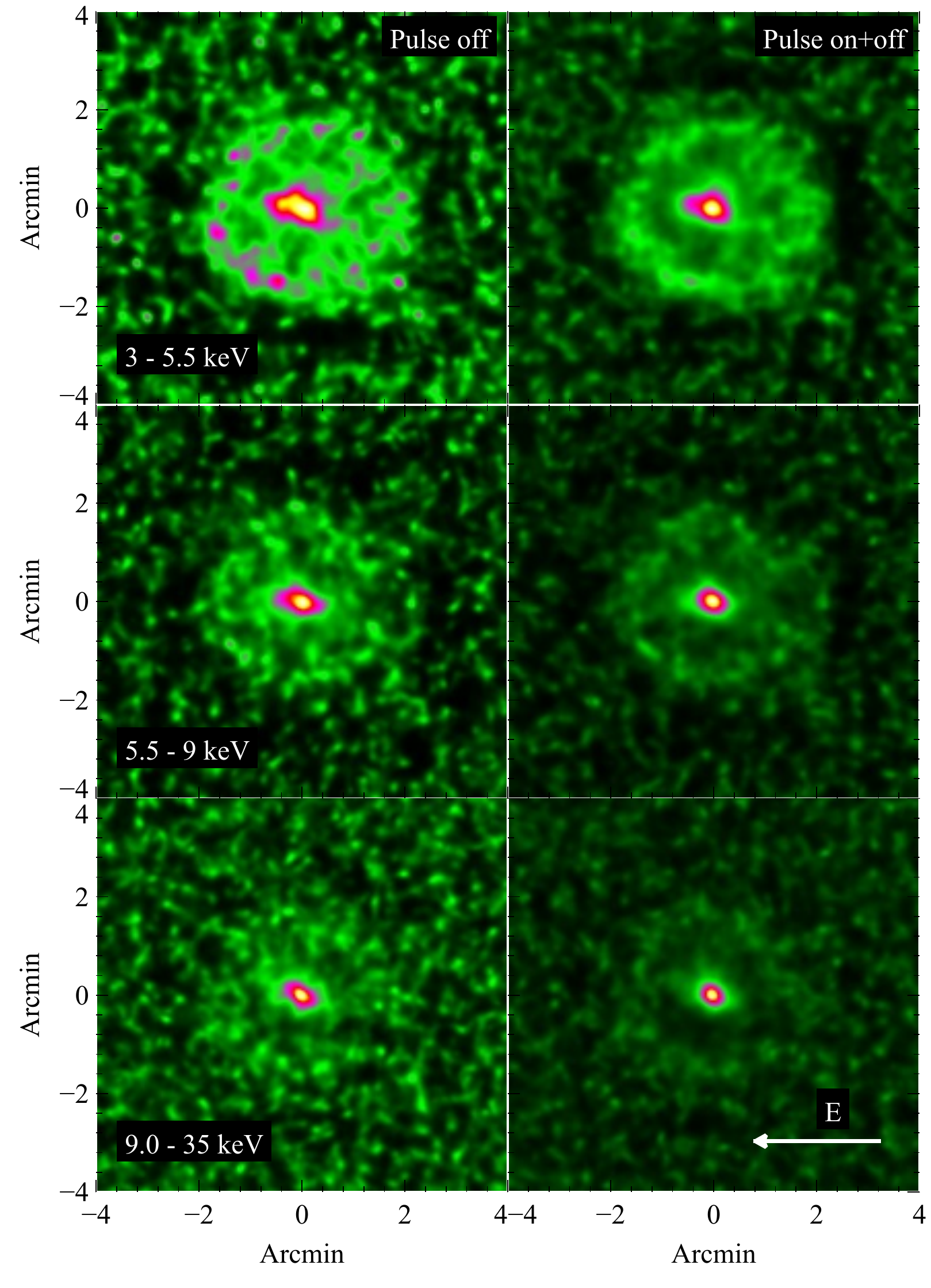}
\end{center}
\caption{Deconvolved images. Left column shows off-pulse and right column pulse on+off (entire phase). From top to bottom the energy bands are: 3 -- 5.5, 5.5 -- 9, and 9 -- 35\,keV.}
\label{deconvolvedimages}
\end{figure}

\subsection{Imaging}\label{sec:imaging}
To investigate the energy-dependent remnant geometry, we first divided each FPM into three energy bands chosen to have equal amounts of counts after background subtraction. The background was obtained adjacent to the remnant, RA=272.9379 and Dec=-19.4457, clear of any straylight or source contamination, and at radius of 80\as\ was kept as large as possible without crossing any detector borders. Having the same number of counts in each band  ensures that the deconvolved images can still be compared even though the flux is not being strictly conserved by the deconvolution process. These energy bands are: 3.0 -- 5.5, 5.5 -- 9.0, and 9.0 -- 35\,keV. We then made two selections, one for the off-pulse (phase 0.6 -- 1.0) and one for the entire phase (0.0 -- 1.0). Each image and module were separately deconvolved using the \textit{max\_likelihood} IDL routine from AstroLib\footnote{See https://github.com/wlandsman/IDLAstro}, which is a maximum-likelihood algorithm based on \citet{Richardson1972} and \citet{Lucy1974}. The deconvolution procedure requires the average background to be zero, and we subtracted the average background from the image. We used the energy appropriate 2D PSFs from the \nustar\ CALDB library and chose the off-axis angle to be the average for the observation. The algorithm requires iterative steps and we used 50 deconvolutions. This number was obtained as optimal by measuring the PSFs of deconvolved point sources and finding that no further improvement was obtained in PSF width beyond this value (for details see \citet{Madsen2015}). After the deconvolution we combined FPMA and FPMB images. 

\begin{figure}
\begin{center}
\includegraphics[width=0.40\textwidth]{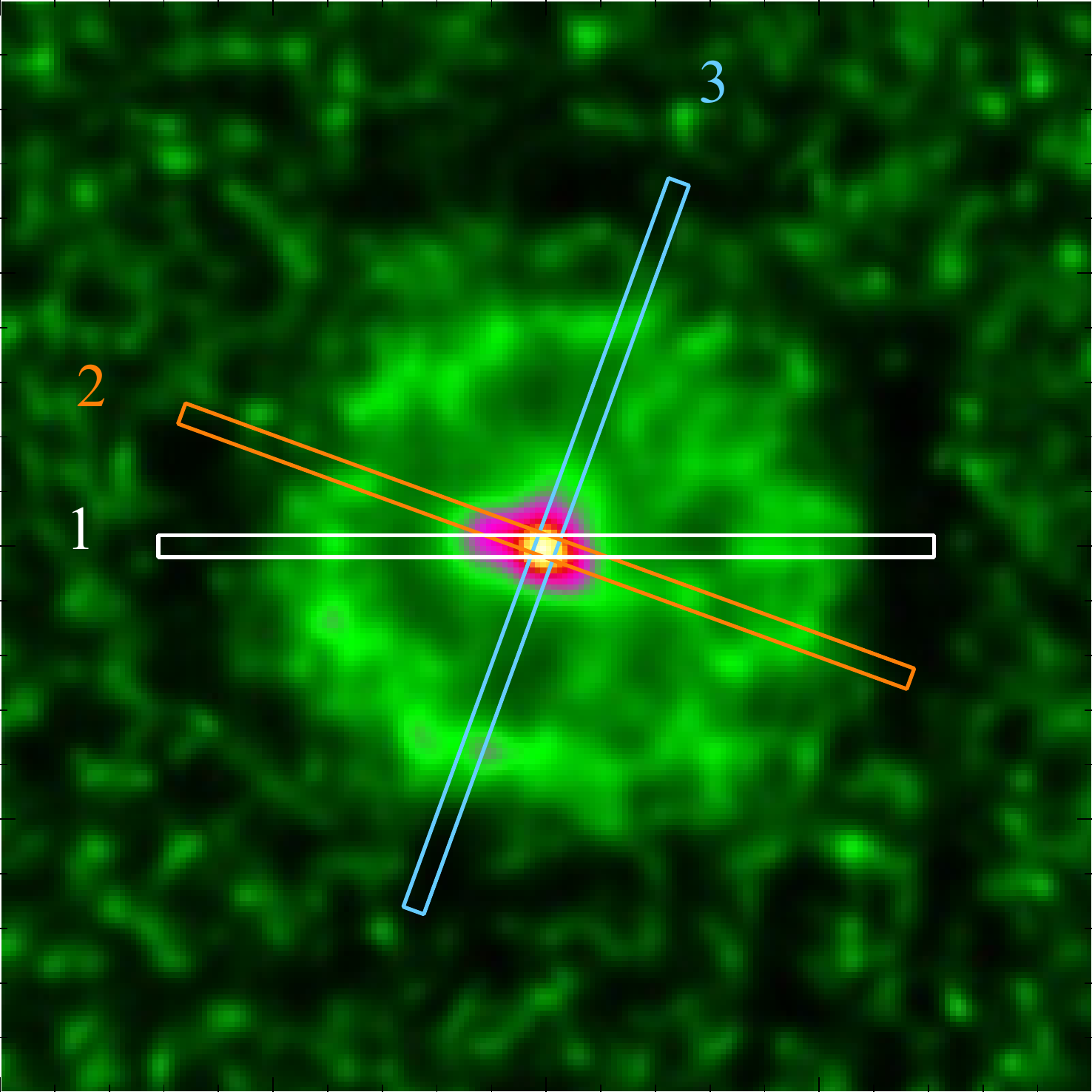}
\end{center}
\caption{The location of the extracted strips 350\as\ long and 17\as\ across: 1) along RA, 2) along the jet-axis, and 3) along the torus axis, perpendicular to the jet.}
\label{stripgeometry}
\end{figure}

\begin{figure}
\begin{center}
\includegraphics[width=0.45\textwidth]{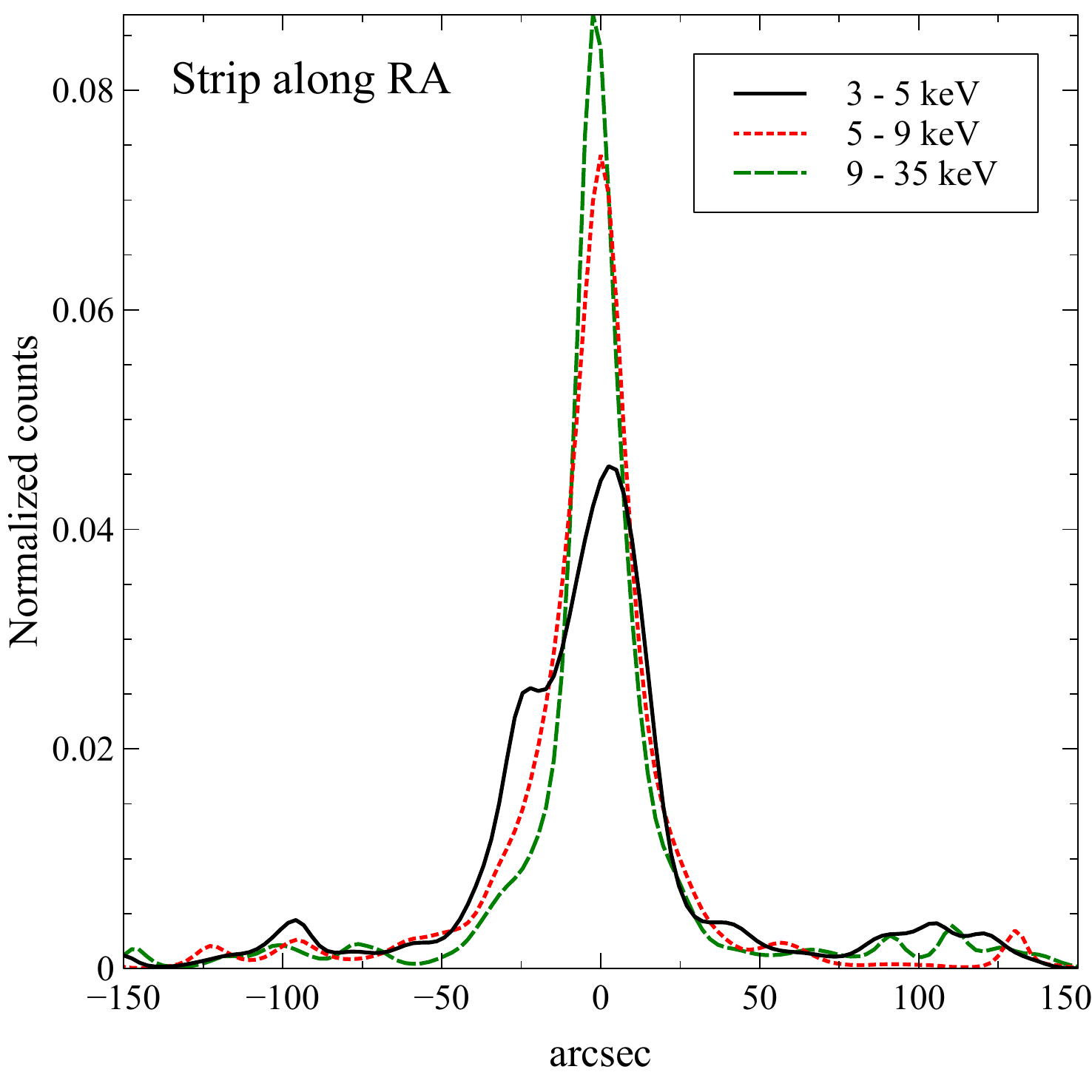}
\end{center}
\caption{Strip extracted along RA 350\as\ long and 17\as\ across (see Figure \ref{stripgeometry}) through the pulsar position and summed across the short axis. The strips have been normalized to the total number of counts in the strip and shows that the intensity becomes more concentrated at the center of the remnant with increasing energy.}
\label{intensityprofile}
\end{figure}

Figure \ref{deconvolvedimages} shows the three energy bands for the two selections. Since each image contains the same amount of photons, and the stretch of each image is the same from peak to background, the apparent dimming of the remnant with respect to the central PWN is real. There is also an indication in the off-pulse images that the PWN is shrinking with increasing energy. To investigate this in more detail, we extracted a 350\as\ strip along RA with a width of 17\as\ in Declination through the pulsar (see Figure \ref{stripgeometry}) and summed across the short axis.  Figure \ref{intensityprofile} shows the intensity profile strips normalized to the total number of counts present in the strip in a linear plot to emphasize how the intensity across the remnant becomes more centralized with increasing energy. It also shows that the peak of the intensity is shifting east. In \chandra, \citet{Borkowski2016} finds the maximum intensity of the jet below 8~keV to be located West of the pulsar location, which is in agreement with our findings. Above 8~keV we see the intensity on the east side of the pulsar decreasing and the maximum intensity localized around the pulsar. We note that since the pulse profile broadens at higher energies, it is possible that there is some pulsar contamination present at the edges of the off-pulse window that could bias the peak intensity towards the pulsar location. 

\begin{figure*}
\begin{center}
\includegraphics[width=0.8\textwidth]{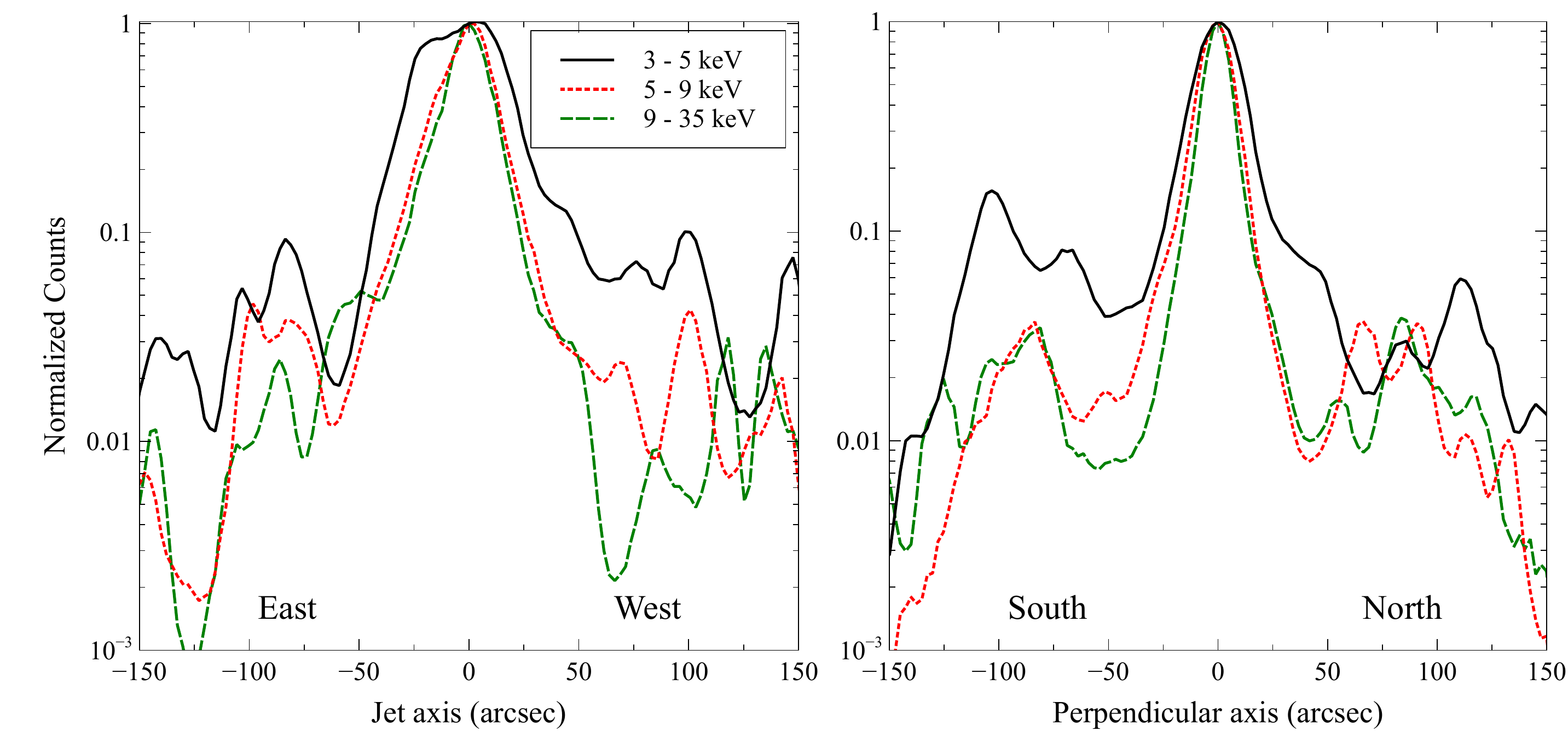}
\end{center}
\caption{Intensity profiles through the pulsar location along the jet axis and perpendicular to the jet axis. The images were rotated by 340\degree and a strip 350\as\ long and 17\as\ across.}
\label{jetprofile}
\end{figure*}

To determine the shrinkage rate in the remnant as a function of energy, we defined two axis; one along the `jet axis' (estimated from \chandra\ images to be 340\degree), and one that is perpendicular, which we call the `torus axis'. We rotated the images by 340\degree\ and again extracted an intensity strip 350\as\ long and 17\as\ wide (see Figure \ref{stripgeometry}) from each energy band and plot them together in Figure \ref{jetprofile}, where we have this time normalized each profile at the pulsar location, which we identify as the geometrical center of the remnant. It should be noted that the maximum intensity is not found at the pulsar location, but off-center along the jet axis as already discussed. We measure the Half Width at Half Max (HWHM) from the pulsar in arcseconds.

\begin{figure}
\begin{center}
\includegraphics[width=0.45\textwidth]{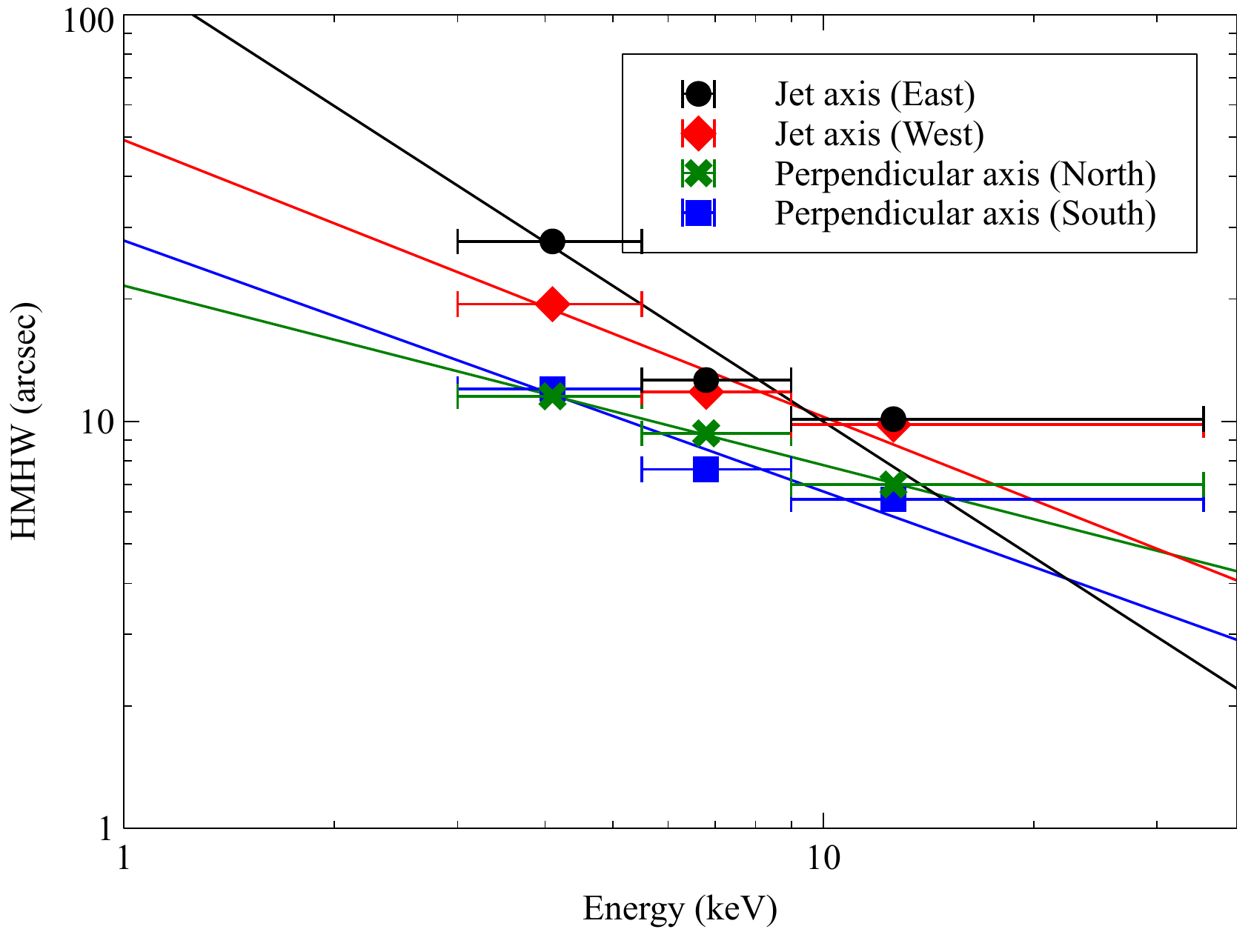}
\end{center}
\caption{Half width half max (HWHM) values measured from the pulsar location.}
\label{hwhm}
\end{figure}

Because the deconvolution procedure does not offer an absolute error on how well it has managed to recreate the correct lengthscale, we deconvolved several strong point sources in the same energy band and noted that their PSF after deconvolution changed by about 1.5\as\ between the lowest and highest band. This gives us a conservative 2\as\ relative error between energy bands. We weighted the HWHM bins with the number of counts for an asymmetric center of the bin, and fitted the HWHM as a function of energy with a power-law: $kE^{-\gamma}$ (see Figure  \ref{hwhm}). The averaged East and West exponent for the jet is  $\gamma_\mathrm{jet} = 0.9 \pm 0.3$, and for the torus axis $\gamma_{torus} = 0.5 \pm 0.4$. Unfortunately, the uncertainties on the function are quite large, but we can still deduce that the central parts of the remnant appear to shrink faster along the jet axis (East-side side more rapidly than the West) than the torus axis (South-side more rapidly than the North).
 
\subsection{Spectroscopy}
With \chandra, \citet{Roberts2003} characterized and measured the spectrum across the remnant using a plane-parallel shock model \citep{Borkowski2001} and a  power-law to account for a hard excess. They observed significant variations across the remnant, however, the results suffered from a partial degeneracy between the absorbing column, the electron plasma temperature, and the power-law index. \nustar\ has little sensitivity to the absorbing column and the abundances in the remnant, but with its wider bandpass it can constrain the power-law index. To overcome these shortcomings in both instruments, we supplement the \nustar\ data with \chandra\ where it benefits.

We use \texttt{XSPEC} for fitting \citep{Arnaud1996}, Wilms abundances \citep{Wilms2000}, Verner cross-sections \citep{Verner1996}, and C-stat as the fitting statistic \citep{Cash1979} on the un-binned data, but report the goodness of fit for the \nustar\ spectra only, since , as shall be explained, it was not feasible to do so for the combined \chandra-\nustar\ fit. Unless otherwise stated the errors will be reported at the 90\% confidence limit.

\subsubsection{Full remnant broadband spectrum}\label{sec:broad}
While \src\ is a complex object with at least three distinct components (PSR, PWN, and shell) we first describe a fit to the spatially integrated emission for comparison with non-imaging instruments such as \textit{NICER} and \textit{INTEGRAL}. We extracted the broadband spectrum from both observatories for the entire remnant within 123\as, including both the PSR and PWN. The PSR is marginally piled up in \chandra, but the skew to the spectrum is minor enough that qualitative assumptions about the spectral shape can still be made if we ignore energies above 5\,keV for \chandra. In \nustar\  we have signal up to 35\,keV, but since the background becomes comparable to the source at $\sim 25$\, keV we fit conservatively from 3 -- 20 \,keV.

\begin{figure}
\begin{center}
\includegraphics[width=0.45\textwidth]{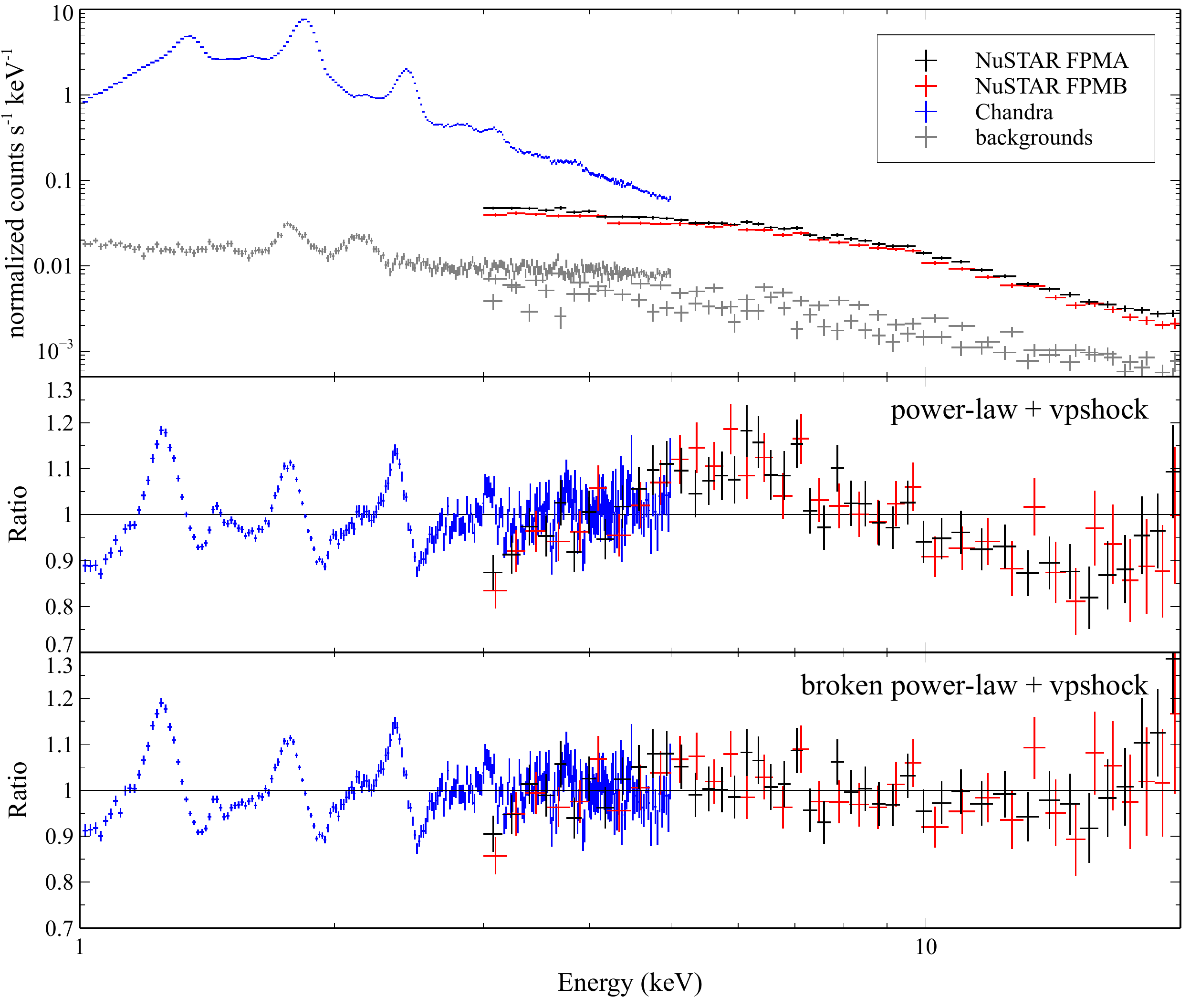}
\end{center}
\caption{Fit to \chandra\ and \nustar\ for the entire remnant, PSR + PWN. Residuals are shown for the two models and it can be seen that a broken power-law greatly improves the residuals, though not perfectly. We interpret the remaining residuals as the inefficacy of the sharp broken power-law cusp to model a gradual break.}
\label{spec1}
\end{figure}

We model the spectra with an absorbed\footnote{http://pulsar.sternwarte.uni-erlangen.de/wilms/research/tbabs/} power-law and a plane-parallel shock model, given in XSPEC notation as: \texttt{tbabs(powerlaw+vpshock)}. We found the upper limit of the ionization timescale, $\tau_u$, to be degenerative with the electron plasma temperature and froze it to $\tau_u= 4.2 \times 10^{11}$ s cm$^{-3}$ as found by \citet{Roberts2003}. Choosing a $\tau_u$ that is 50\% lower increases the electron temperature by $\sim 10\%$. Separately, each instrument fits this model well, but as already discussed, the \nustar\ data offers poor constraints on the shock component, while the \chandra\ data poorly constrains the power-law component. The best fit values therefore, unsurprisingly, differed significantly between instruments. However, when fitted together, we discovered that this discrepancy isn't just due to the different energy bands, but because there are significant residuals in the transition region between the thermal and non-thermal component between 5 -- 7\,keV as shown in Figure \ref{spec1}, middle panel, which appears like a 'break' in the continuum. 

As reported by \citet{Madsen2017b}, cross-calibration differences are known to exist between the two observatories, mainly in the absolute normalization, but also in the slopes of the two instruments. However, it is important to note that the feature is seen only in \nustar\ well outside the \chandra\ band and not directly in the overlap region, which is where typically cross-calibration issues show themselves. Unfortunately, dedicated campaigns between \nustar\ and \chandra\ only exist for the gratings and we can therefore not rule out that this feature could stem from a cross-calibration issue. There are strong indications, though, that this is not the case. A $\sim9$~keV break was detected in G21.5-0.9 with \nustar\ \citep{Nynka2014}, and breaks across the Crab PWN were seen between 8 -- 12~keV also exclusively with \nustar\ data \citep{Madsen2015}. In MSH~15-52 a break was detected around 6.3~keV \citep{An2014}, and like for \src, this one was only found when combined with \chandra\ data. Outside of PWN spectra, we have not observed breaks like this between \nustar\ and \chandra\ data, and we therefore proceeded under the assumption that this is not a cross-calibration issue, but an actual problem with the chosen models.
  
A better fit was obtained by replacing the power-law with a broken power-law and setting the break energy around $\sim 6$~keV (see Figure \ref{spec1}, bottom panel). This improved model still exhibits residuals around 5 -- 6~keV, where we introduced the break, but they may be explained  by a far more gradual transition than the sharp cusp of the broken power-law.  As to the source of the break, one possibility is that the thermal component is poorly represented by a single shock and rather should be a superposition of several different electron temperatures and abundances. This is supported by the very large residuals seen in our abundance lines of Mg, Si, S, Ar and Ca. \citet{Lopez2011} measured abundances at 23 locations across the remnant and found the abundances and electron temperatures can vary by factors of 2. Adding more \texttt{vpshock} components indicated that these residuals could be improved, but it introduced a large number of free parameters and degeneracy that raised its own complications. We therefore continued with the broken power-law interpretation, noting that the power-law index, $\Gamma_1$, below the break energy has little physical meaning.

In the initial modeling we allowed the abundances to remain free and find relative to solar: Mg$\sim$1.2, Si$\sim$1.5, S$\sim$1.3, Ar$\sim$1.2, and Ca$\sim$2.9. These abundances are acceptably close to average abundance values inferred from \citet{Lopez2011}, but as stated the fit leaves significant residuals in the lines. These residuals can be cosmetically improved upon by adding a number of gaussians, but since we already understand the reason for the residuals, and the application of multiple gaussians is unphysical, we left the residuals as they are. 

At this point we therefore emphasize that the inclusion of the \chandra\ data serves to better define the power-law by providing constraints on the thermal component and the galactic absorption. It is not our intent to make any detailed measurements or statements on the abundances values, which has already been covered in detail by \citet{Roberts2003}, \citet{Lopez2011}, and \citet{Borkowski2016}. Instead, we focus here on the hard non-thermal component, represented by $\Gamma_2$, which is the index above the power-law break.

We proceeded to freeze the abundances and calculated errors for all relevant parameters, noting that it became necessary to evaluate the electron temperature from the $\chi^2$-curve produced by the steppar command due to the complications from the line residuals. For the entire remnant, including the PSR, we find the best fit parameters: \nh=3.35$\pm 0.01$, $kT=0.65\pm 0.01$,  $E_\mathrm{break}=5.9\pm0.3$, $\Gamma_1=1.2 \pm 0.1$, and $\Gamma_2= 1.78 \pm 0.03$ (see Table \ref{spectraltable}). 

\begin{figure}
\begin{center}
\includegraphics[width=0.4\textwidth]{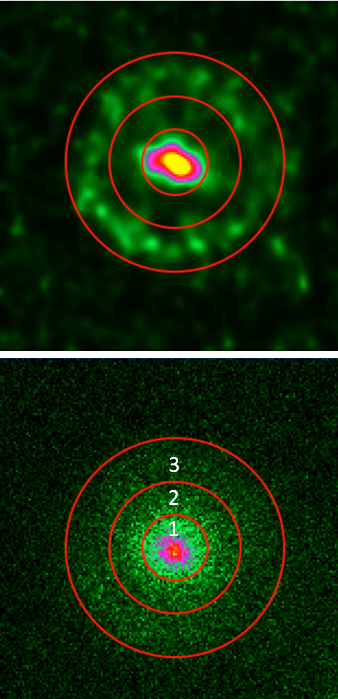}
\end{center}
\caption{Top: The deconvolved 3 -- 35\,keV image. Bottom: The raw image of FPMA. Red circles correspond to radius of: (\#1) 37\as, (\#2) 74\as, and (\#3) 123\as.}
\label{extractionannuli}
\end{figure}

\begin{table*}
\centering
\caption{Spectral fits}
\begin{tabular}{|l|c|c|c|c|c|c|c|c|c|}
\hline 
\texttt{model} (XSPEC)  & \texttt{tbabs} & \multicolumn{3}{c|}{\texttt{broken powerlaw}} & \multicolumn{2}{c|}{\texttt{logpar}\footnote{Alternative to \texttt{powerlaw} used for pulsar only}} & \multicolumn{2}{c|}{\texttt{vpshock}} & \\
\hline
parameter &  \nh (tbabs)\footnote{unit: 10$^{22}$ atoms cm$^{-2}$} & $E_{break}$ & $\Gamma_1$ & $\Gamma_2$ & $\alpha$ & $\beta$ & kT & $\tau_u$ & $\bar{\chi}^2$ \\
& & & &&&& (keV) & ($10^{11}$s\,cm$^{-3}$) & ($\chi^2$/dof)\footnote{The Goodness of Fit is reported for the \nustar\ data only (with \chandra\ data removed).} \\
\hline
remnant all\footnote{Includes pulsar and nebula for all phases for an extraction region of radius 123\as} & 3.35$\pm 0.01$ & $5.9\pm0.3$ & $1.2 \pm 0.1$ & $1.78 \pm 0.03$ & - & - & 0.65$\pm 0.01$ & 4.2\footnote{Parameter frozen} & 1394/1348 \\
PWN region 1 (0 -- 37\as) & 3.21$\pm 0.03$ & $3.5 \pm 0.3$ & $0.6 \pm 0.3$ & $1.66 \pm 0.05$ & - & - & 0.62$\pm 0.2$& $4.2^\mathrm{e}$ &  131/140\\
PWN region 2 (37 -- 74\as) & 3.29$\pm 0.02$ & $4.9 \pm 0.3$ & $0.3 \pm 0.4$ & $1.92 \pm 0.07$ & - & - &  0.63$\pm 0.01$ & 4.2$^\mathrm{e}$ & 198/166 \\
shell region 3 (74 -- 123\as) & 3.49$\pm 0.01$ & $5.5 \pm 0.2$ & $0.2 \pm 0.4$ & $2.1 \pm 0.1$ & - & - & $0.64 \pm 0.01$ & 4.2$^\mathrm{e}$ & 438/399\\
\hline
 &  & \multicolumn{3}{c|}{\texttt{powerlaw}} & \multicolumn{2}{c|}{} & \multicolumn{2}{c|}{} & \\
\hline
pulsar\footnote{Pulse on - pulse off} & 3.2$^\mathrm{e}$ &&  $1.32 \pm 0.07$ & - & - & - & - & - & 650/641 \\
pulsar$^\mathrm{f}$& 3.2$^\mathrm{e}$ & - & - & - & $1.14 \pm 0.2$ & $0.31 \pm 0.29$ & - & - & 647/640\\
\hline
\end{tabular}
\label{spectraltable}
\end{table*}

\begin{table*}
\centering
\caption{Spectral fits: \nustar\ only}
\begin{tabular}{|l|c|c|c|c|c|}
\hline 
\texttt{model} (XSPEC)  & \texttt{tbabs} & \texttt{powerlaw}  & \multicolumn{2}{c|}{\texttt{vpshock}} & \\
\hline
parameter &  \nh (tbabs)  & $\Gamma$ &  kT & $\tau_u$ & $\bar{\chi}^2$ \\
       &10$^{22}$ atoms cm$^{-2}$&& (keV) & ($10^{11}$s\,cm$^{-3}$) & ($\chi^2$/dof) \\
\hline
PWN region 1 (0 -- 37\as)   & 3.2\footnote{Parameter is frozen} & 1.86$\pm 0.05$  & -- & -- & 139/153 \\
PWN region 2 (37 -- 74\as) & 3.3$^\mathrm{a}$ & 1.96$\pm 0.13$ & 0.6$^{+0.4}_{-0.2}$ & 4.2$^\mathrm{a}$ & 193/158 \\
Shell region 3 (74 -- 123\as) & 3.5$^\mathrm{a}$ & 2.2$\pm 0.1$ & 0.5$^{+0.1}_{-0.1}$ & 4.2$^\mathrm{a}$ & 323/305 \\
\hline
\end{tabular}
\label{tablenustaronly}
\end{table*}


\begin{figure}
\begin{center}
\includegraphics[width=0.45\textwidth]{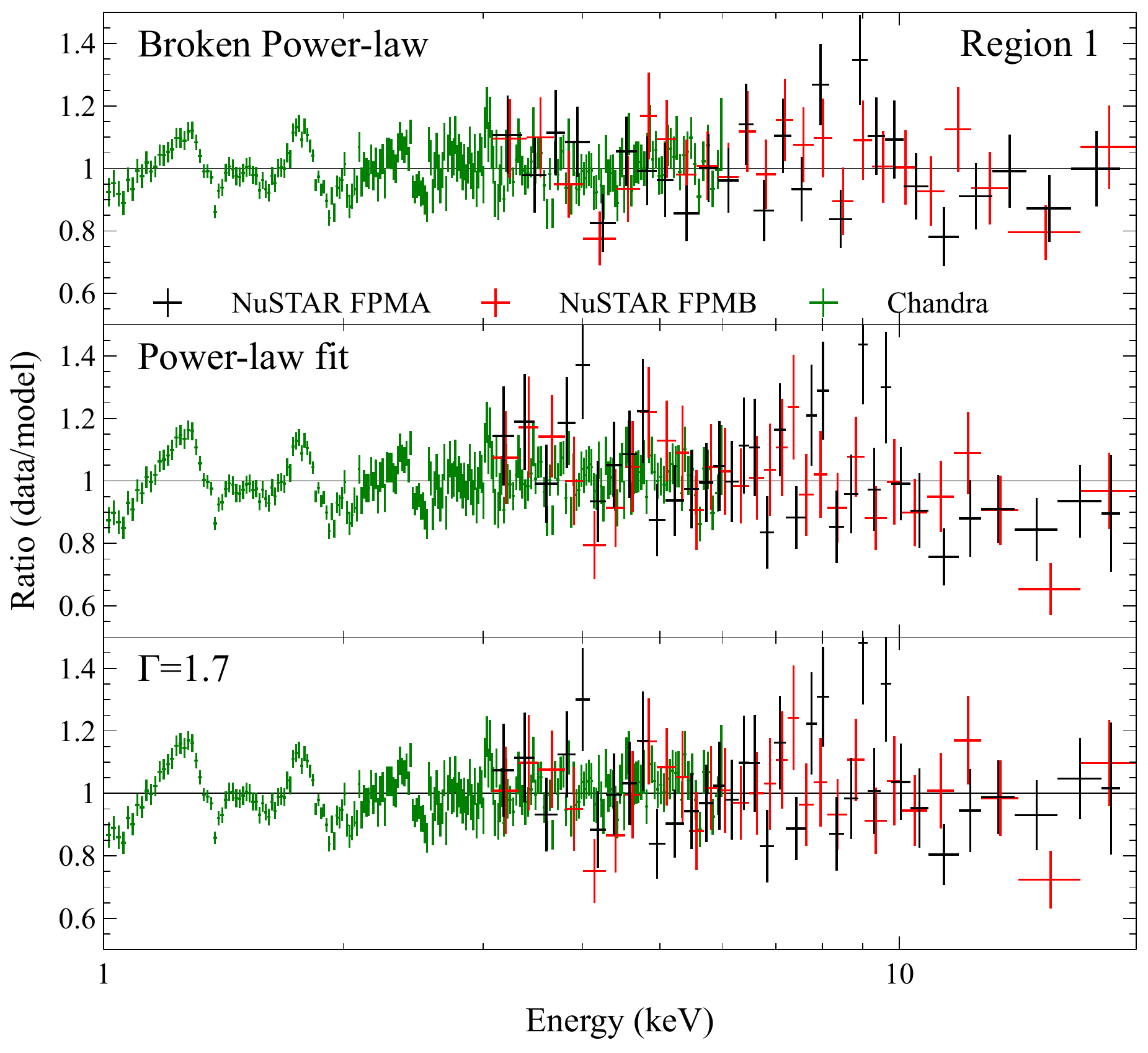}
\end{center}
\caption{Combined fits to \chandra\ and \nustar\ for region \#1 shown in Figure \ref{extractionannuli}. Top panel are residuals to the best fit of a broken power-law and a plane parallel shock. Middle panel are residuals to a power-law and plane parallel shock. Bottom panel are residuals then the power-law index is frozen to $\Gamma_2$ given in Table \ref{spectraltable}. }
\label{spec2}
\end{figure}

\begin{figure}
\begin{center}
\includegraphics[width=0.45\textwidth]{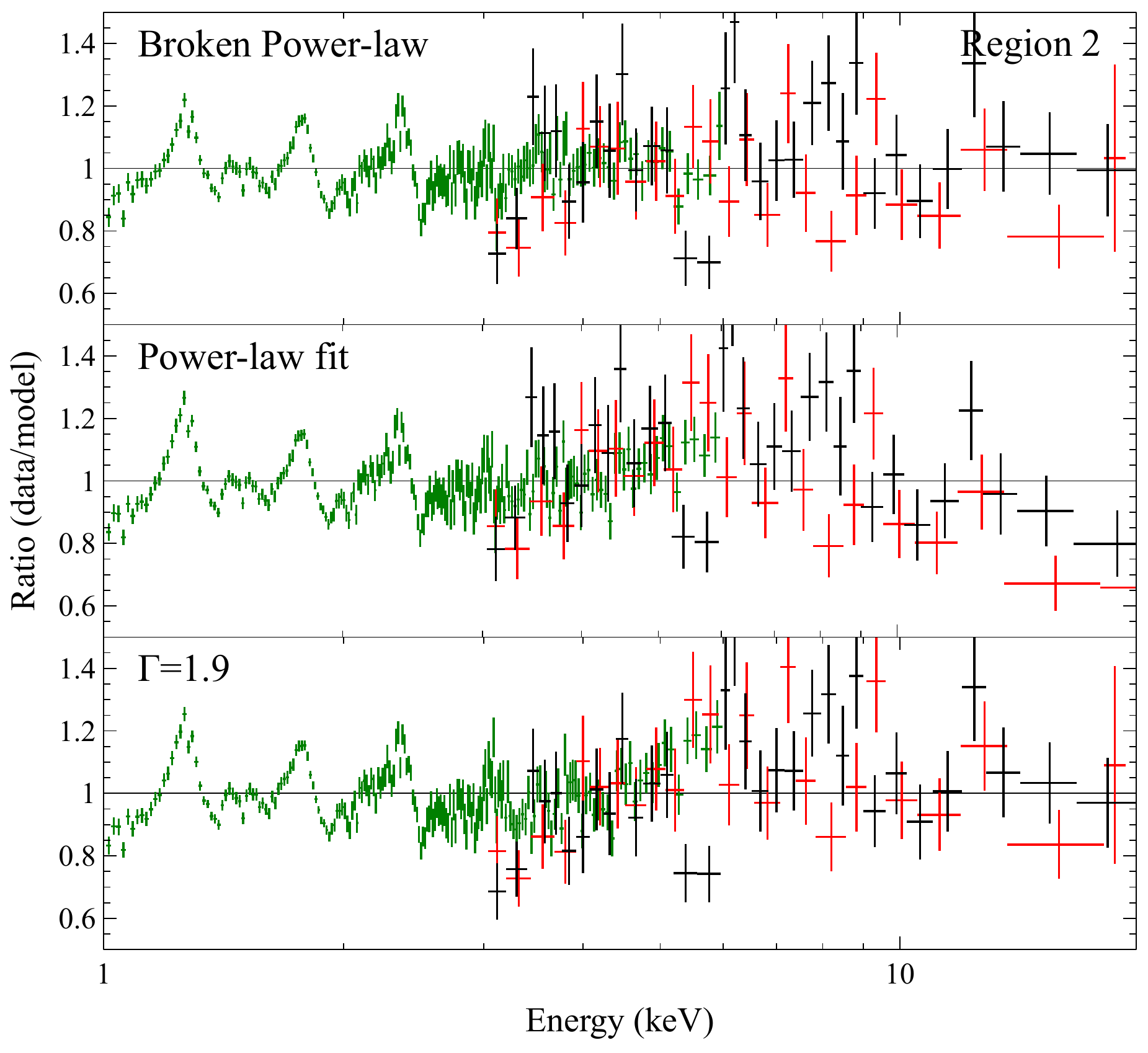}
\end{center}
\caption{Combined fits to \chandra\ and \nustar\ for region \#2 shown in Figure \ref{extractionannuli}. Top panel are residuals to the best fit of a broken power-law and a plane parallel shock. Middle panel are residuals to a power-law and plane parallel shock. Bottom panel are residuals then the power-law index is frozen to $\Gamma_2$ given in Table \ref{spectraltable}.}
\label{spec3}
\end{figure}

\begin{figure}
\begin{center}
\includegraphics[width=0.45\textwidth]{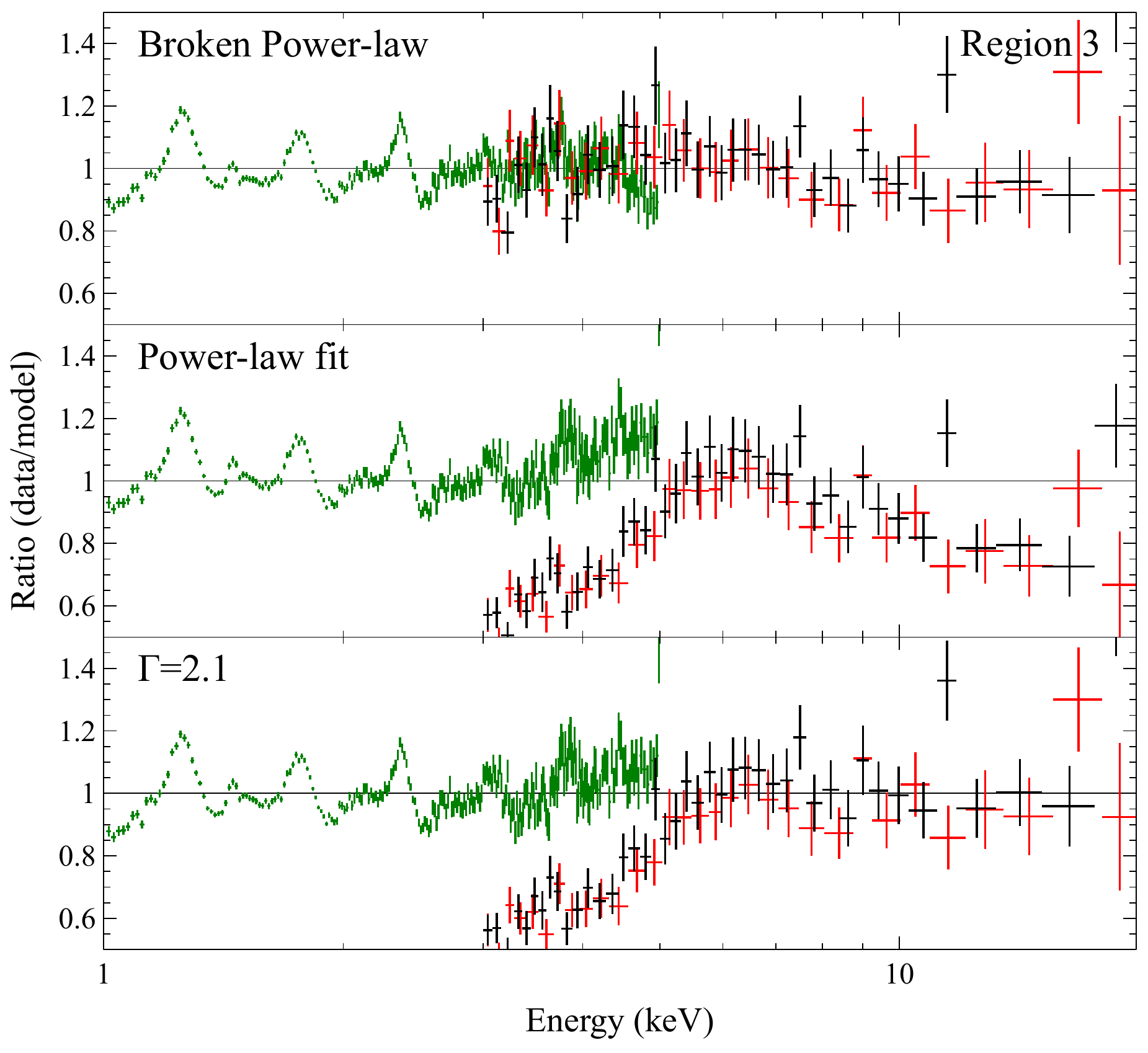}
\end{center}
\caption{Combined fits to \chandra\ and \nustar\ for region \#3 shown in Figure \ref{extractionannuli}. Top panel are residuals to the best fit of a broken power-law and a plane parallel shock. Middle panel are residuals to a power-law and plane parallel shock. Bottom panel are residuals then the power-law index is frozen to $\Gamma_2$ given in Table \ref{spectraltable}.}
\label{spec4}
\end{figure}

\subsubsection{Nebula and Shell Spectrum}\label{sec:nebula}
To investigate the PWN spectrum, we separated out the PSR from the nebula in the \nustar\ data by employing phase-resolved spectroscopy. This is necessary due to fact that the \nustar\ PSF, which has a Half Power Diameter (HPD) of $~60$\as, will otherwise contaminate the remnant with the PSR. We used the ephemeris presented in section \ref{sec:pulseprofile} to extract the counts from phase bins 0.6 -- 1.0 (see Figure \ref{pulseprofile}), which we define as the off-pulse phase. As shown in Figure \ref{extractionannuli}, we extracted three regions for both instruments: 0 -- 37\as (\#1), 37 -- 74\as(\#2), and 74 -- 123\as(\#3), which cover the central PWN, the shell, and what is between. Because the shell, \# 3, has very little contamination from the pulsar, confirmed by the absence of pulsations when folding the spectrum on the period and verifying identical fits within errors to the pulse-on and -off spectrum, we used the full phase range (0.0 -- 1.0) to increase statistics. For \nustar\ the response files are obtained out of the standard pipeline for extended regions. Since PSF corrections cannot be applied to extended responses in \nustar, the flux will not be precise between instruments, and we allowed for a constant to account for the flux differences between FPMA and FPMB and another for \chandra.    

Based on our findings above, we used the broken power-law model, \texttt{bknpowerlaw + vpshock}, and found that in all three cases it is required. The line residuals in \chandra\ still dominate the fit statistics and we followed the procedure outlined before and froze the abundances once the residuals had been minimized. We summarize the fit results in Table \ref{spectraltable}, and find that $\Gamma_2$ softens with increasing radius, progressing from region the PWN ( \#1) through to the shell (\#3) from $1.66\pm 0.05$ to $2.1 \pm 0.1$, which supports the scenario of the remnant becoming fainter with increasing radius and energy. 

As a separate check, we fitted the \nustar\ data alone, which requited us to freeze the \nh\ and abundances to the value found with \chandra\ for each region and the upper ionization timescale to $\tau_u=4.2 \times 10^{11}$s\,cm$^{-3}$, neither of which are sensitive, or can be constrained, in the \nustar\ band. The fits to the \nustar\ data alone are given in Table \ref{tablenustaronly} and show that within errors the photon index is consistent with that found for $\Gamma_2$ in Table \ref{spectraltable}. For the PWN (region \#1) we cannot measure a thermal component in \nustar.

To quantify the broken power-law spectrum better, we also fitted the combined spectra with a power-law in place of the broken power-law and did one fit with the power-law photon index left free, and another fixing the photon index to $\Gamma_2$ from the broken power-law found in the same region. The ratio plots for all three regions are shown in Figure \ref{spec2}, \ref{spec3}, and \ref{spec4}. In the upper panels we show the broken power-law fit, in the middle panel the power-law fit, the results of which we do not record since they are for visual purposes only, and in the bottom panel the power-law fit with $\Gamma=\Gamma_2$ frozen. By considering the ratios in each region, it appears that the broken power-law spectrum is strongest in the shell. 

Finally, we fitted the spectra in the shell with a \texttt{vpshock+srcut} model, where \texttt{srcut} describes the synchrotron spectrum from an exponentially cut-off power-law distribution of electrons in a homogeneous magnetic field. We used the values for the radio spectral index, $\alpha=0.56$, and normalization of the radio flux at 1 GHz of 2 Jy, obtained from \citet{Tam2002}, but the curvature of the spectrum at high energies is far too quick to describe the data.  

We searched for an iron line in the \nustar\ data, but find no evidence of its presence. Despite background and pileup issues around the iron region, we also searched the \chandra\ data since a line should still be evident even if the continuum is piled-up, but do not find any evidence of iron there either.

\subsubsection{Pulsed spectrum}\label{sec:pulse}
From the imaging analysis it is apparent that the PWN contributes less flux at higher energies, which is supported by the pulse fraction curve, showing the ratio of the PSR flux to PWN flux increasing as a function of energy. To investigate the shape of the pulsed spectrum, we used as background the off-pulse phase (0.6 -- 1.0) and subtracted it from the on-pulse phase (0.0 -- 0.6). To test the stability of the results, we used three different extraction regions: the entire remnant (123\as), an intermediate region (74\as), and the interior (37\as). They agree within errors, so we used the highest SNR spectrum from a radius of 74\as. Because the background is contained in the same region, which reduces the uncertainties, we were able to measure the spectrum all the way up to 50\,keV. We fitted the spectrum with two models: a \texttt{powerlaw} and \texttt{logpar} model, which is a power-law model where the photon index varies as a log parabola 
\begin{equation}
F(E) = K (E/E_1)−^{(\alpha+\beta log(E/E_1 ))} \mathrm{ph\,cm}^{-2}\mathrm{\,s}^{-1}\mathrm{\,keV}^{-1}\,.
\end{equation}
Here $\alpha$ is the photon index at the pivot energy $E_1$, and we set $E_1 = 5$\,keV. The results of the two models are shown in Table \ref{spectraltable}, and while the \texttt{logpar} model yields a slightly better fit, the difference between the two models is only significant to 98\%.

\subsection{Broadband SED}\label{sec:sed}
Armed with this understanding of the PWN and PSR, we then proceeded to fit the spectrum of the entire remnant, PWN+PSR, (here we include the shell together with the PWN) across \chandra, \nustar, and \integral, from 1  --  300\,keV. To help with the stability of the spectrum and to illustrate how the different components interact, we included four spectra from \nustar: the full phase (phase: 0 -- 1.0), on-pulse period (phase: 0.0 -- 0.6), off-pulse period (phase: 0.6 -- 1.0), and the pulse on-off spectrum. 

We fit with the model  \texttt{vpshock + bknpowerlaw(PWN)+powerlaw(PSR)} and set the normalization of the PSR to 0 in the off-pulse spectrum, and the normalization of the PWN (thermal and non-thermal) to 0 for pulse on-off spectrum. As before we freeze the abundances in \texttt{vpshock} and set $\tau_u=4.2 \times 10^{11}$s\,cm$^{-3}$. The resulting fit is good with a $\bar{\chi}^2$=1.08 (1392/1285) for a $\Gamma_\mathrm{PWN}=2.01 \pm 0.08$, $\Gamma_\mathrm{PSR}=1.34 \pm 0.08$, and $kT = 0.75 \pm 0.08$~keV, summarized in Table \ref{broadbandtable}. 

\begin{figure}
\begin{center}
\includegraphics[width=0.45\textwidth]{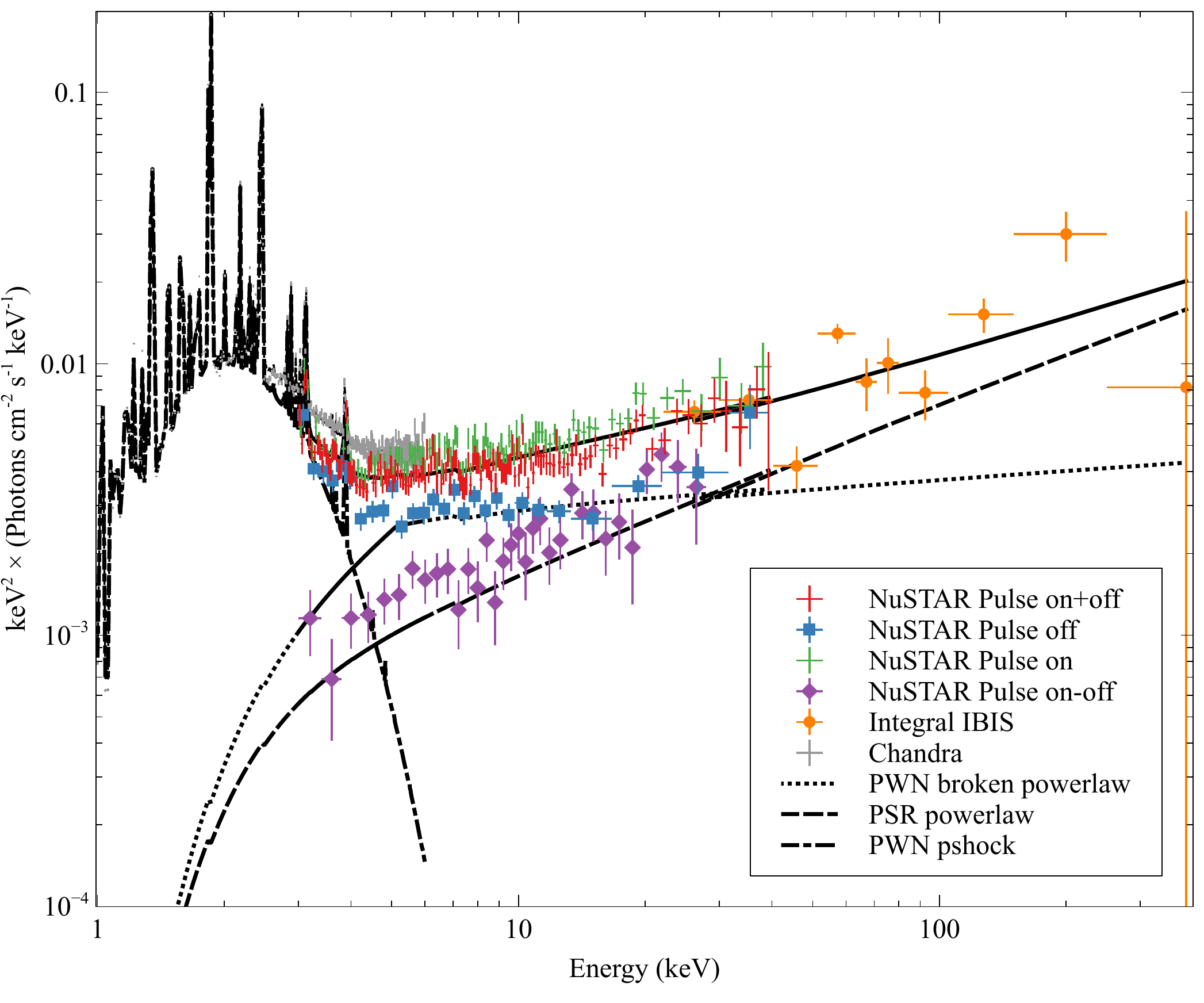}
\end{center}
\caption{Broadband SED using \chandra, \integral, and \nustar. }
\label{SED3}
\end{figure}

If we compare these results to those obtained in Table \ref{spectraltable}, it is reassuring that when adding in the \integral\ data we recover the same result. Figure \ref{SED3} shows $\nu F_\nu$ and illustrates that the hard X-ray spectrum is composed of the two non-thermal components, one from the PWN + shell and the other from the PSR, which grows to dominate above 20\,keV. To break this down even further, we can calculate the flux in 5 -- 20\,keV from the \nustar\ data in the three regions and find that the PWN and shell contribute roughly equally. The harder PWN ($\Gamma \sim 1.7$) will eventually dominate over the shell ($\Gamma \sim 2.1$) with increasing energy, but neither will have any considerable contribution at $\gamma$-ray energies compared to the PSR.


\begin{table*}
\centering
\caption{1 -- 300\,keV Broadband Spectral Fit}
\begin{tabular}{|l|c|c|c|c|c|c|c|}
\hline 
\texttt{model} (XSPEC)  & \texttt{tbabs} & \multicolumn{2}{c|}{\texttt{powerlaw} (PSR)}  & \multicolumn{2}{c|}{\texttt{powerlaw} (PWN)} & \texttt{vpshock}\footnote{\,$\tau_u=4.2 \times 10^{11}$s\,cm$^{-3}$, abundances relative to solar: Mg = 1.1, Si=1.4, S=1.2, Ar=1.1, Ca=2.7.} & \\
\hline
parameter &  \nh\footnote{\,10$^{22}$ atoms cm$^{-2}$}  & $\Gamma$ &  N\footnote{\,photons\,keV$^{-1}\mathrm{cm}^{-2}\mathrm{s}^{-1}$} & $\Gamma$ & N$^\mathrm{c}$ & kT (keV) & $\chi^2$/dof \\
\hline
 & $3.5 \pm 0.01 $ & $1.34 \pm 0.08$  & $3.9 \pm 0.8 \times 10^{-4}$  & $2.01 \pm 0.08$ & $3.6 \pm 0.8 \times 10^{-4}$ & $0.75 \pm 0.08$ & 1392/1285\footnote{\,With \chandra\ data removed} \\
\hline
\end{tabular}
\label{broadbandtable}
\end{table*}

\section{Discussion}

We can collect our findings in three categories: the PSR, PWN, and the shell.  

For the PSR we find 
$P=6.4706254242 \times 10^{-2}$~s and $\dot P=3.4332573 \times 10^{-14}$~s~s$^{-1}$ and if we assume the magnetic dipole field of a canonical pulsar with $R=10$\,km and moment of inertia of $I=10^{45}$g\,cm$^{-2}$ this gives a minimum field strength and spin down rotational energy of   
\begin{eqnarray}
B &>& \left ( 3 c^3 I \over 8 \pi^2 R^6 \right)^{1/2}(P\dot P)^{1/2} = 1.5 \times 10^{12} \mathrm{G}, \\
\dot E &=& \left ( 4 \pi^2 I \dot P \over P^3 \right ) = 5.0 \times 10^{36} \mathrm{ergs\,s}^{-1}.
\end{eqnarray}

We find a rise in pulsed fraction with energy, which is explained by the underlying PWN continuum contributing less flux at higher energies with respect to the pulsar. We find that the pulsed spectrum can be described by a power-law with photon index $\Gamma=1.35 \pm 0.08$ all the way through the \integral/IBIS band up to 300~keV.  At 20~keV the contribution of PWN+shell and PSR to the total flux is roughly 50/50, but at higher energies, the pulsed spectrum dominates the combined flux of the remnant and becomes the primary contributor in the soft $\gamma$-rays. The pulsed spectrum is consistent with the measurement made with \rxte\ reported by \citet{Roberts2004}, and the interpretation of the pulsed flux dominating above 20\,keV consistent with previous \rxte\ and \integral\ findings by \citet{Kuiper2015}.


It is commonly accepted that the sources of the high-energy radiation in rotation powered pulsars are curvature and synchrotron photons from pair production cascades in the magnetosphere. However, the site of the acceleration has long been a matter of debate, and though it still remains uncertain, \fermi\ re-solved the long-standing question of whether the acceleration originated close to the stellar surface from the polar cap region \citep{Daugherty1982} or in the outer magnetosphere at the light cylinder (where the velocity of the co-rotating magnetic field equals the speed of light), as in the outer-gap region \citep{Cheng1986}  and the slot-gap region along the current layers at the boundary between closed and open field lines \citep{Arons1983}.
Magnetic pair production in the strong fields above the polar cap predicts steep, super-exponential absorption cut-offs in the $\gamma$-ray spectra above a few GeV, which have not been observed. Instead, \fermi\ detected pulsars typically exhibit hard photon spectra with a gradual decline at several tens of GeV \citep{Abdo2010, Abdo2013}. Despite being young and bright in the X-rays, \psr\ has not been detected in the \fermi\ band. The flux of the power-law extrapolated into the \fermi/LAT band (100 MeV -- 100 GeV) is $\sim1 \times 10^{-5}$ photons cm$^{-2}$ s$^{-1}$, which is well above the detection threshold limit of $1 \times 10^{-9}$ photons cm$^{-2}$ s$^{-1}$ for a photon index of 1.5 given by Figure 20 in \citet{Abdo2010}. This would indicate that the spectrum has a turnover below 100~MeV and makes it similar to PSR J1846--0258 \citep{Kuiper2009} also detected with \integral\ ISGRI/IBIS but not by \fermi/LAT, and PSR B1509−-58, which is detected above 1 MeV with a measured cutoff of a few MeV \citep{Cusumano2001, Abdo2010, Pilia2010}. Common for these three is that they all have broad, single pulsed profiles in contrast to the typical narrow double-peaked $\gamma$-ray pulsars. Recent progress in Particle-in-cell (PIC) simulations of pulsar magnetospheres indicate that the likely location of high-energy particle acceleration occurs along the current sheet at the equator in a zone close to the light-cylinder \citep{Chen2014,Philippov2015,Philippov2018}. In the context of these simulations, the characteristics of \psr\ can be understood if it has an inclined magnetic axis in the range 30--60\degree\ and is viewed at an angle of $\sim$45\degree. In this case, one observes a single pulse peak originating from the electron populations with a spectral energy distribution that falls off faster than for double peaked pulse profiles \citep[see Figure 10,][]{Cerutti2016}. Detailed analysis performed with radio and X-rays, which take into account the shape of the torus and asymmetric brightness of the jets, indicates that the tilt of the torus to the plane of the sky is $\sim 60$\degree \citep{Borkowski2016}. The line of sight to the rotation axis is then 30\degree\ and consistent with what can be inferred from the pulse profile and cut-off of the spectrum.

The spatially integrated spectrum of the pulsar-wind nebula at energies above the {\sl Chandra} band is well described by a power-law with $\Gamma = 1.71 \pm 0.07$, consistent with the value of $1.78 \pm 0.7$ reported by \cite{Borkowski2016}.  That study found no significant steepening in spectrum with distance from the pulsar as is seen in other PWNe such as G21.5$-$0.9 \citep{Nynka2014}; while the nominal best-fit values of $\Gamma$ did increase, the magnitude of the change was within errors.  However, here we find that the PWN extent along the jet direction shrinks with increasing energy, with HWHM $\propto E^{-\gamma}$ with $\gamma_{jet} = 0.9 \pm 0.3$ along the jet axis and $\gamma_{torus} =0.5 \pm 0.4$ perpendicular to that direction.  While errors are large, the shrinkage along the jet seems secure and larger than that perpendicular to the jet.  These two results are consistent if particle transport along the PWN is primarily advective and monotonically increasing going out, in which case one expects a constant spectrum until an abrupt spectral cutoff at a distance from the pulsar corresponding to the particle lifetimes \citep[see figures in][]{Reynolds2003}. Most PWNe show, instead, gradual steepening of the spectrum, indicating a mixture of particles of different ages at a given distance from the pulsar, such as might be produced by diffusion \citep[e.g.,][]{Reynolds91,Tang2012} or more complex advective motions \citep{Porth2013}.  The magnitude of the energy-dependent shrinkage exponent, $\gamma$, of about $0.9$ is distinct from those measured by {\sl NuSTAR} in G21.5$-$0.9 \citep{Nynka2014} and MSH 15$-$5{\sl 2} \citep{An2014}, where values of $\gamma$ of about $0.2$ were found.  For the Crab, \cite{Madsen2015} reported differing shrinkage rates along the jet, counterjet, and transverse (torus) directions of about $0.05$, $0.2$, and $0.08$, respectively.  Thus the jet in \src\ stands out among PWNe in two ways: a more rapid shrinkage with energy, but absence of progressive spectral steepening along its length.  Evidently the nature of particle transport in \src\ is different from that in other very young PWNe.  A deeper analysis of the PWN in Kes 75, the youngest known in the Galaxy, may cast light on this situation.

For the shell we confirm the suggestions from {\sl Chandra} data that a non-thermal component is required, and can be described by a power-law with $\Gamma = 2.1 \pm 0.1$.  An {\tt srcut} fit does not do well in describing the data.  We also see some shrinkage of the shell radius with increasing energy, consistent with the {\sl Chandra} finding that harder emission is concentrated near the inner edge of the shell \citep{Borkowski2016}.  

The confirmation of non-thermal X-rays from the shell means that
\src\ joins the other Galactic remnants less than a few thousand years
old in having evidence for shock acceleration of electrons to
multi-TeV energies.  Only three Galactic shell remnants of
core-collapse supernovae with ages less than about 2000 yr are known:
Cas A (about 350 years old), Kes 75 \citep[about $480 \pm 50$ years
  old;][]{Reynolds18}, and \src. The non-thermal X-ray spectrum of
Cas A is remarkable, extending as a single power-law to energies of
order 100 keV, with the hardest emission originating from neither the
forward nor the reverse shock \citep{Grefenstette2015}.  While the {\sl
  Chandra} spectrum of the shell in Kes 75 requires a hard spectral
component, that component may be a power-law \citep{Helfand2003} or a
high-temperature thermal component from the blast wave
\citep{Morton2007}.  All remnants of Type Ia supernovae from the last
2000 years (G1.9+0.3, Tycho, Kepler, SN 1006, and RCW 86; see
\cite{Reynolds08} for a review) show synchrotron X-ray emission of
unambiguous character.

The shell spectrum in \src\ is quite hard compared to synchrotron X-ray
emission from the other shells.  For Cas A, $\Gamma \sim 3.1$ for
filaments associated with the forward shock, and $3.4$ for interior
emission above 15 keV \citep{Grefenstette2015}.  For the young Type Ia
remnants, $\Gamma \sim 3$ is typical \citep[ e.g., $\Gamma = 3.0$ for Tycho's SNR;][]{wang14}
Our value of $\Gamma = 2.1$ for the photon index corresponds to an
energy index $\alpha_x$ of 1.1, about 0.5 larger than the radio energy
index of $\alpha = 0.56$.  Of all the known cases of shell synchrotron
X-rays, only G11.2-0.3 shows such a small amount of steepening.  The
value $\Delta \alpha = 0.5$ is of course the expectation for the very
simple case of continuous electron acceleration to very
high energies followed by radiative losses in a homogeneous source.
Synchrotron losses simultaneous with acceleration will produce an
electron spectrum with an (approximately) exponential cutoff at an
energy at which the acceleration time equals the loss time \citep{zirakashvili07}, rather than a steeper power-law.  The sum of a range
of cut-off spectra can produce a power-law.

An attempt to model the integrated spectral-energy distribution (SED)
of the shell emission of \src\ with a simple loss model encounters
severe quantitative difficulties.  If we take our observed value of
$\Gamma$ at face value, the extrapolations of the radio spectrum
($S_\nu \sim 20 (\nu/1 \ {\rm GHz})^{-0.6}$) up and the X-ray spectrum
down meet at a frequency of about $3 \times 10^{12}$ Hz.  If we
picture electrons as accelerated in a region in which the magnetic
field allows energies of $\approx 100$ TeV to be reached, but then
radiating subsequently in a region with higher field strength, our
knowledge of the age of \src\ of about 2000 years allows the deduction
of that higher magnetic-field strength.  The half-life $t_{1/2}$ of an
electron radiating the peak of its synchrotron spectrum at frequency
$\nu$ in a magnetic field $B$ is given by
\begin{equation}
  t_{1/2} = 5.69 \times 10^{11}\,B^{-3/2}\,\nu^{-1/2}\ {\rm s}.
\end{equation}
For an age of 2000 years and a break frequency of $3 \times 10^{12}$ Hz,
we find
\begin{equation}
  B = 300\ \left(t \over 2000\ {\rm y}\right)^{-2/3} \,
  \left(\nu \over 3 \times 10^{12}\ {\rm Hz}\right)^{-1/3}\ {\mu \rm G}.
\end{equation}
This value is quite high, perhaps implausibly so.

This naive picture is almost certainly incorrect.  The maximum photon
energy emitted by an electron distribution limited by radiative losses
depends only on the shock velocity \citep[and geometric factors likely
  to be of order unity; e.g.,][]{Reynolds08}: $h\nu_{\rm max, loss}
\sim 0.2 u_8^2$ keV where $u_8$ is the shock speed in units of $10^8$
cm s$^{-1}$.  For \src, proper-motion observations give $u_8 \sim (0.7
- 1.2)$ \citep{Borkowski2016}, so it is impossible to produce the
observed 20 keV synchrotron photons from an electron distribution
accelerated in a region with a magnetic-field strength of 300 $\mu$G.


But the interpretation of the integrated SED of \src\ as that of a
power-law steepened by continuous losses already requires that the
conditions in the acceleration region be different from those in the
regions where the electrons do most of their radiating.  To produce
synchrotron photons up to the $\sim 20$ keV we observe from the
radiating region where $B \sim 300\ \mu$G, we require electron
energies up to $E_m \sim 60$ TeV.  The loss-limited maximum electron
energy is roughly $E_{\rm m, loss} \sim 100 u_8 B_{\mu{\rm
    G}}^{-1/2}$ TeV; since $u_8 \sim 1$, we require a very low
magnetic field in the acceleration region, of order 1 $\mu$G.  The
combination of a very low field near the shock, where electrons are
presumably accelerated, followed by their diffusing or advecting into
a region where $B$ is larger by orders of magnitude, seems extremely
implausible.  Much more likely is that the X-ray photon index reflects
not the spectrum radiated by a single power-law electron distribution,
but the superposition of distributions accelerated under a range of
conditions, with the rough agreement of $\alpha_r + 0.5$ with the
X-ray energy index $\Gamma - 1$ entirely fortuitous.

\section{Conclusions}

Our {\sl NuSTAR} observations extend the range of X-ray studies of \src\ to 35 keV, with new results on the pulsar, the pulsar-wind nebula, and the outer shell.  The PSR shows a pulse profile broadening with increasing energy, and an increasing pulsed fraction, and its spectrum does not show evidence of curvature up to 300~keV. The PWN has an integrated spectrum consistent with earlier studies, but shows shrinkage along the jet direction, which contrasts with the lack of observed spectral steepening along the jet in {\sl Chandra} observations.  The electron outflow in the PWN may be simpler than that seen in other young PWNe.  Our imaging observations of the shell show a slightly smaller radius at higher energies, consistent with {\sl Chandra} results.  We confirm the existence of a hard, power-law component from the shell of \src, with photon index $\Gamma = 2.1 \pm 0.1$, which is almost certainly synchrotron emission, given the absence of significant Fe K$\alpha$ emission between 6.4 and 6.7 keV. While this value of $\Gamma$ agrees with the expected value for the index of synchrotron emission from a simple model of synchrotron losses in a homogeneous source given the radio (energy) spectral index of 0.6, the implied ``break'' frequency of $3 \times 10^{12}$ Hz demands an impossibly high magnetic field, and the agreement is likely fortuitous.  Instead, we attribute the hard spectrum to a superposition of spectra from electrons accelerated in different regions with different conditions, which may also explain the broken spectrum of the power-law.

{\it Facilities:} \facility{NuSTAR} \facility{Chandra} \facility{INTEGRAL}

\acknowledgments This work was supported under NASA Contract No. NNG08FD60C, and made use of data from the \textit{NuSTAR} mission, a project led by the California Institute of Technology, managed by the Jet Propulsion Laboratory, and funded by the National Aeronautics and Space Administration. We thank the \textit{NuSTAR} Operations, Software and Calibration teams for support with the execution and analysis of these observations. This research has made use of the \textit{NuSTAR} Data Analysis Software (NuSTARDAS) jointly developed by the ASI Science Data Center (ASDC, Italy) and the California Institute of Technology (USA). We would also like to thank Robert Archibald for his help with finding the timing solution.

\bibliography{bib}
\bibliographystyle{jwaabib}

\end{document}